\documentclass[journal, onecolumn]{IEEEtran}
\usepackage{cite,amsmath}       
\IEEEoverridecommandlockouts

\usepackage{graphicx}
\usepackage{multirow}

\usepackage{amssymb}
\usepackage{tabulary}
\usepackage{booktabs}
\usepackage{amsmath}
\usepackage{subfigure}
\usepackage{epsfig}
\usepackage{sidecap}
\usepackage{wrapfig}
\usepackage{stfloats}
\usepackage{array}
\usepackage{mathtools}
\usepackage{cite}
\usepackage{url}
\usepackage[mathscr]{euscript}
\usepackage{arydshln}
\usepackage{subfigure}
\usepackage{color}
\usepackage{tabulary}
\usepackage{enumitem}
\usepackage{algorithm}
\usepackage{algpseudocode}
\usepackage{optidef}
\usepackage{microtype}
\allowdisplaybreaks
\usepackage{parskip}
\usepackage[papersize={8.5in,11in}, left=0.7in, right=0.7in, top=0.7in, bottom=0.8in]{geometry}

\algnewcommand\algorithmicforeach{\textbf{for each}}
\algdef{S}[FOR]{ForEach}[1]{\algorithmicforeach\ #1\ \algorithmicdo}

\newcommand{\bigO}{\mathcal{O}}

\begin{document}

\title{\huge{An Alternating Algorithm for Uplink Max-Min SINR  in  Cell-Free Massive MIMO with  Local-MMSE Receiver}}
\author{W. A. Chamalee Wickrama Arachchi, K. B. Shashika Manosha, Nandana Rajatheva, and M.\ Latva-aho
        \\Centre for Wireless Communications, University of Oulu, Finland.
        \\\{\small{chamalee.wickramaArachchi@student.,  manosha.kapuruhamybadalge@, nandana.rajatheva@, matti.latva-aho@\}oulu.fi}}
\maketitle
\thispagestyle{plain}
\pagestyle{plain}
\begin{abstract}

The problem of max-min signal-to-interference plus noise ratio (SINR)  for uplink transmission of cell-free massive multiple-input multiple-output (MIMO) system is considered. We assume that the system is employed with local minimum mean square error (L-MMSE) combining. The objective is to preserve user fairness by solving max-min SINR optimization problem, by optimizing transmit power  of each user equipment (UE)  and weighting coefficients at central processing unit (CPU), subject to transmit power constraints of UEs.
This problem is not jointly convex. Hence, we decompose original problem into two subproblems, particularly for optimizing power allocation and receiver weighting coefficients.
Then, we propose an alternating algorithm to solve these two subproblems. The weighting coefficient subproblem is formulated as a generalized eigenvalue problem while  power allocation subproblem is approximated as  geometric programming (GP). We empirically show that the proposed algorithm achieves higher min-user uplink spectral efficiency (SE) over existing fixed  power scheme which is not optimized with respect to the transmit power. Moreover, the convergence of the proposed algorithm is numerically illustrated.
\end{abstract}


\section{Introduction}
Massive MIMO (mMIMO) has been recognized as an emerging technology towards fifth generation (5G) wireless networks and beyond due to dramatically improved spectral efficiency (SE)  without sacrificing additional bandwidth and transmit power  resources~\cite{5595728}. The key idea of mMIMO is to deploy base stations~(BS) with large antenna arrays which serve many user terminals simultaneously, in the same time-frequency resource exploiting  spatial degrees of freedom. However, co-location of many antennas together may lead to correlation in the channel. To cope with this  major  realization challenge in mMIMO, distributed antenna system (DAS) has been proposed~\cite{viability_2013}. The basic idea of DAS is that multiple antennas of
a BS are geographically separated within the cell. Prior work in~\cite{viability_2013} shows  that DAS is capable of increasing average rates compared with a centralized solution. Subsequently, cooperative DAS, also referred  as cooperative multiple point~(CoMP) is another promising way to improve SE further through BS cooperation~\cite{comp,comp_distributed_2010}. 

Cell-free  mMIMO has been proposed in~\cite{cellfreemMimoVsSmallCells} as an incarnation of various notions of MIMO, particularly ``network MIMO'', ``distributed MIMO'', ``DAS'', ``CoMP'', and
etc. Cell-free  mMIMO setup is same  as cooperative DAS; but with no cell-boundaries. 
Prior work in~\cite{DBLP:journals/corr/abs-1804-03421} illustrates that cell-free mMIMO is capable of providing more uniform performance among users by co-processing over access points~(APs). This is due to the fact that, in contrast to conventional network-centric cellular networks, cell-free mMIMO provides a user-centric implementation to overcome inter-cell interference. Furthermore, two key properties; favorable propagation and increased macro diversity are numerically illustrated in~\cite{DBLP:journals/corr/abs-1804-03421} to explain this paradigm shift of cell-free mMIMO compared with cellular networks. To elaborate these two properties further, favorable propagation offers  near orthogonality between channel vectors of any pair of UEs and hence, users are capable of sharing same time-frequency resource without incurring high inter-user interference.  Moreover, facilitated by the distributed architecture of cell-free mMIMO, macro-diversity is now increased due to the fact that each UE is served by muliple APs rather than single BS.
In addition, the excessive handover issue in small-cell systems can be solved using cell-free topology. Thus, cell-free mMIMO has attracted a lots of research interest recently.

Unlike in conventional cellular networks, cell-free mMIMO system deploys large number of distributed APs over a geographical area where number of users are much lower than number of APs. In the uplink transmission, all users  simultaneously transmit in the same time-frequency resource block to all APs and each AP performs multiplexing with linear receiver processing techniques. In the existing literature,  maximum ratio (MR), regularized zero-forcing (RZF), local minimum mean-squared error (L-MMSE) and other MMSE processing techniques are considered in ~\cite{8422577,bjrnson2019newRZF,DBLP:journals/corr/abs-1903-10611}. More specifically, the study in~\cite{DBLP:journals/corr/abs-1903-10611} shows that higher spectral efficiency is achieved by cell-free  mMIMO when
using L-MMSE combining  compared with MR processing. To ensure user fairness, max-min power control is utilized in the  cell-free networks~\cite{cellfreemMimoVsSmallCells}. This mechanism provides fairness to all users, irrespective of their geographical location. 
 Moreover, power control is carried out at CPU on the large-scale fading time scale~\cite{cellfreemMimoVsSmallCells}, since CPU does not have the channel estimates; but only channel statistics. However, utilizing the channel statistics, CPU  improves SE further by optimizing  receiver weighting coefficients~\cite{8422577}. 

 The focus of previous study~\cite[Corollary 2]{DBLP:journals/corr/abs-1903-10611}  was to maximize individual signal-to-interference plus noise ratio (SINR) by only changing weighting coefficients; but, taking transmit powers
as fixed. In contrast to previous work, our goal
is to maximize the minimum SINR of the system by changing
both transmit power and weighting coefficients. In this paper,
we propose an alternating algorithm to solve  max-min SINR
problem for the uplink of a cell-free mMIMO system 
with L-MMSE combining. This max-min SINR problem is not jointly convex. Hence,  we divide the original problem into two subproblems; particularly for power allocation and weighting coefficient design to cater for the non-convexity of original max-min SINR problem. Weighting and power coefficient subproblems are solved by formulating as a generalized eigenvalue problem~\cite{Golub:1996:MC:248979} and approximating as  geometric programming~(GP)~\cite{boyd_vandenberghe_2004} respectively. We extend proposed optimization methodology to a variety of cell-free mMIMO setups. Numerical results show that the proposed algorithm achieves higher minimum user uplink SE in comparison with the SE obtained by using fixed transmit power in~\cite[Corollary 2]{DBLP:journals/corr/abs-1903-10611}. In addition, we present the computational complexity of the proposed algorithm.


The rest of the paper is organized as follows: Section~\ref{sec:system-model} describes the system model for uplink cell-free mMIMO which employs L-MMSE detector. Next, Section~\ref{Sec:prob_formulation} presents the proposed algorithm to solve max-min SINR optimization problem. In Section~\ref{sec:complexity}, we discuss the computational complexity of the proposed algorithm. Section~\ref{sec:results} numerically illustrates  the effectiveness of the proposed algorithm. Finally, Section~\ref{sec:conclusion} concludes this paper.

\subsection{Notation}
Boldface lowercase and uppercase letters denote vectors
and matrices, respectively, and calligraphy letters denote sets. The superscripts $^{T}$, $^*$ and $^{H}$ denote transpose, conjugate, and conjugate transpose, respectively. 
 The multi-variate circularly symmetric complex Gaussian distribution with correlation matrix $\mathbf{R}$ is denoted by $\mathcal{N}_{\mathbb{C}}(\mathbf{0}, \mathbf{R})$ whereas $ \mathcal{CN}(0, \sigma^2)$  stands for circularly symmetric complex Gaussian distribution with zero mean, variance $\sigma^2$, and $ \mathcal{N} (0, \sigma^2 )$ denotes real-valued Gaussian distribution. 
 The set of complex n dimensional vectors is denoted by  $\mathbb{C}^{n}$ and the set of complex
$m \times n$ matrices is denoted by $\mathbb{C}^{m \times n}$. The expected value of $\mathbf{x}$ is denoted as $\mathbb{E}\{ \mathbf{x} \}$.
The $n \times n$ identity matrix is represented as $\mathbf{I}_n$. Finally, the absolute value of the complex number x is denoted by $|x|$ and Euclidean norm of the vector $\mathbf{x}$ is denoted by $\| x \|$.  

\section {System model}
\label{sec:system-model}
\begin{figure}[t!]
\center
\includegraphics[width=76mm]{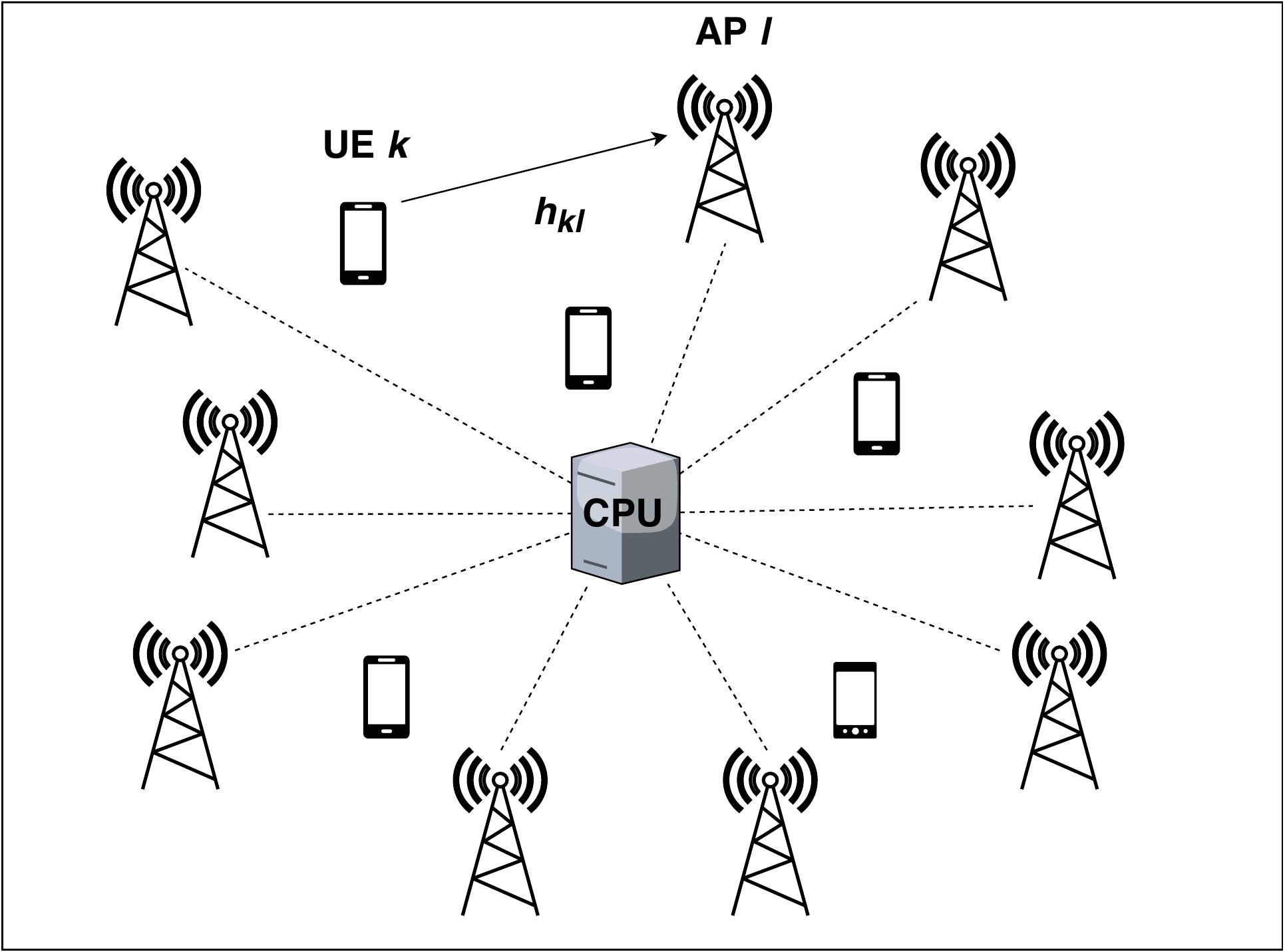}
\caption{The uplink of a cell-free Massive MIMO system with \textit{K} users and \textit{L} APs. }
\label{sysmodel}
\end{figure}

We consider  a cell-free mMIMO  system with $L$ APs each equipped with $N$ antennas and $K$ single antenna users randomly distributed in a large area, as shown in Fig. \ref{sysmodel}. We assume that $K\ll L$. Each AP is connected to the central processing unit (CPU) via a fronthaul connection. Let $\mathbf{h}_{kl}$ be the channel coefficient vector between $l$th AP  and $k$th UE where $\mathbf{h}_{kl} \in \mathbb{C}^N$. We model the channel using  block fading model where $\mathbf{h}_{kl}$ is fixed during  time-frequency blocks of $\tau_c$ samples. In other words, coherence block consists of $\tau_c$ number of samples. Channel coefficients are independent and identically distributed (i.i.d) random variables. In each block, $\mathbf{h}_{kl}$ is  an independent realization from a  correlated Rayleigh fading distribution defined as follows:
\begin{equation}
\mathbf{h}_{kl} \sim \mathcal{N}_{\mathbb{C}}(\mathbf{0}, \mathbf{R}_{kl}),
\end{equation}
where $\mathbf{R}_{kl} \in \mathbb{C}^{N \times N}$ is the spatial correlation matrix.
\subsection{Pilot Transmission and Channel Estimation}

We consider uplink transmission where all users send uplink pilots and payload data in $\tau_p $  
and   $\tau_c - \tau_p$ samples respectively. To estimate the channel coefficients in the uplink, all users send  pilot sequences of length  $\tau_p$ simultaneously to all the APs. 
\\
Let $\tau_p$ number of mutually orthogonal  pilot signals  $\boldsymbol{\phi}_{1},\ldots,\boldsymbol{\phi}_{\tau_p}$ with $\| \boldsymbol{\phi}_{t} \|^2 = \tau_p$ are used for channel estimation. We consider a large network where $K > \tau_p$ in which each pilot signal is assigned to more than one UE . The pilot index assigned to $k$th UE is represented as $t_k$, where $t_k \in \{ 1, \ldots, \tau_p\}$.
Then, the received pilot matrix $\mathbf{Z}_{l} \in \mathbb{C}^{N \times \tau_p}$ at $l$th AP is given by 
\begin{equation} \label{eq:received-pilot-matrix}
\mathbf{Z}_{l} = \sum_{i=1}^{K} \sqrt{p_i } \mathbf{h}_{il} \boldsymbol{\phi}_{t_i}^{T}+ \mathbf{N}_{l},
\end{equation}
where $p_i \geq 0$ is the transmit power of $i$th UE , $\mathbf{N}_{l}  \in \mathbb{C}^{N \times \tau_p}$ is the  noise at AP. The elements of $\mathbf{N}_{l}$ are assumed to be independent and identically distributed as $ \mathcal{CN} (0, \sigma^2 )$
and  $\sigma^2$ is the noise power.

Upon receiving $\mathbf{Z}_{l}$ at $l$th AP, it  first correlates  $\mathbf{Z}_{l} $ with corresponding normalized pilot signal $\boldsymbol{\phi}_{t_k}/ \sqrt{\tau_p}$. The correlated signal can be simplified as follows: 
\begin{align} \notag
 \mathbf{z}_{t_k l} \triangleq \frac{1}{\sqrt{\tau_p}} \mathbf{Z}_{l}  \boldsymbol{\phi}_{t_k}^{*}
 & = \sum_{i=1}^{K}  \frac{\sqrt{p_i }}{\sqrt{\tau_p}}\mathbf{h}_{il} \boldsymbol{\phi}_{t_i}^{T}  \boldsymbol{\phi}_{t_k}^{*}+ \frac{1}{\sqrt{\tau_p}} \mathbf{N}_{l}  \boldsymbol{\phi}_{t_k}^{*}  \\
 & = \sum_{i \in \mathcal{P}_k} \sqrt{p_i \tau_p } \mathbf{h}_{il} + \mathbf{n}_{t_k l},
 \label{eq:received-pilot}
\end{align}
where $\mathbf{n}_{t_k l} \sim \mathcal{N}_{\mathbb{C}}(\mathbf{0}, \sigma^2 \mathbf{I}_{N})$. After performing correlation operation,
the MMSE estimate of channel coefficient vector between $l$th AP and $k$th UE, $\hat{\mathbf{h}_{kl}}$ can be written as~\cite{DBLP:journals/corr/abs-1903-10611}
\begin{equation} \label{eq:estimates}
\hat{\mathbf{h}}_{kl} = \sqrt{p_k \tau_p} \mathbf{R}_{kl} \mathbf{\Psi}_{t_kl}^{-1} \mathbf{z}_{t_kl},  
\end{equation}
where
$\mathbf{\Psi}_{t_kl} = \mathbb{E} \{ \mathbf{z}_{t_k l} \mathbf{z}_{t_k l}^{H} \} = \sum_{i \in \mathcal{P}_k} \tau_p p_i \mathbf{R}_{il} + \mathbf{I}_{N}$
is the correlation matrix of the received signal $\mathbf{z}_{t_k l} $.
The channel estimate $\hat{\mathbf{h}}_{kl}$ and the channel estimation error $\tilde{\mathbf{h}}_{kl} = \mathbf{h}_{kl} - \hat{\mathbf{h}}_{kl}$ are independent vectors with $\hat{\mathbf{h}}_{kl}\sim \mathcal{N}_{\mathbb{C}} \left( \mathbf{0}, p_k \tau_p \mathbf{R}_{kl} \mathbf{\Psi}_{t_kl}^{-1} \mathbf{R}_{kl} \right)$ and $\tilde{\mathbf{h}}_{kl}\sim \mathcal{N}_{\mathbb{C}}(\mathbf{0},\mathbf{C}_{kl})$ where
\begin{equation}
\mathbf{C}_{kl} = \mathbb{E} \{ \tilde{\mathbf{h}}_{kl} \tilde{\mathbf{h}}_{kl}^{H} \}= \mathbf{R}_{kl} - p_k \tau_p \mathbf{R}_{kl} \mathbf{\Psi}_{t_kl}^{-1} \mathbf{R}_{kl}.
\end{equation}

\subsection{Uplink Payload Data Transmission}
All users send their signals simultaneously to all APs during the uplink payload data transmission phase. We denote the transmit signal from $i$th UE as  ${s}_{i}\sim \mathcal{CN} (0, p_i )$ and $p_i$ is the transmit power at  $i$th UE. 
The received signal at $l$th AP is given by

\begin{equation} \label{eq:received-data}
\mathbf{y}_{l} = \sum_{i=1}^{K} \mathbf{h}_{il} s_i + \mathbf{n}_{l},
\end{equation}
where receiver noise at $l$th AP is denoted as $\mathbf{n}_{l}\sim \mathcal{N}_{\mathbb{C}} (\mathbf{0}, p_i)$.
 
We assume that an L-MMSE detector is employed at each AP and the received signal at  $l$th AP is first pre multiplied by $\mathbf{v}_{kl}$ where $\mathbf{v}_{kl} \in \mathbb{C}^{N}$ is the local combining vector at  $l$th AP to estimate $s_k$. Let $\check{s}_{kl}$ is the local estimate of kth user at lth AP. The local estimate of $s_k$ is given by
\begin{equation} \label{eq:local-data-estimate}
\check{s}_{kl} \triangleq \mathbf{v}_{kl}^{H} \mathbf{y}_{l} = \mathbf{v}_{kl}^{H}\mathbf{h}_{kl} s_k +  \sum_{i=1,i\ne k}^{K} \mathbf{v}_{kl}^{H} \mathbf{h}_{il} s_i + \mathbf{v}_{kl}^{H}\mathbf{n}_{l}.
\end{equation}
The mean squared error (MSE) of $k$th symbol at  $l$th AP is denoted  by ${\rm{MSE}}_{kl} =\mathbb{E} \{ | s_{k} - \mathbf{v}_{kl}^{H} \mathbf{y}_{l} |^2  \big| \{ \hat{\mathbf{h}}_{il} \}  \}$. The optimal combining vector which minimizes the MSE can be derived by obtaining the first derivative of conditional expectation and setting it to zero. Hence, the optimal combining vector which minimizes the MSE is given by   
\begin{equation} \label{eq:MMSE-combining-single-AP}
\mathbf{v}_{kl} =  p_{k}  \left( \sum\limits_{i=1}^{K} p_{i} \left( \hat{\mathbf{h}}_{il} \hat{\mathbf{h}}_{il}^{H} + \mathbf{C}_{il} \right) + \sigma^2  \mathbf{I}_{N} \right)^{-1}     \hat{\mathbf{h}}_{kl}.
\end{equation}

The pre processed signals using combining vectors at each AP are then forwarded to CPU for final signal detection. The forwarded signals are further  multiplied by weighting coefficients  at CPU to improve achievable rate. CPU does not have the knowledge of the channel estimates and therefore, only channel statistics are utilized to maximize SE. 
\\
Let  $a_{kl}$ is the weighting coefficient of $k$th user at $l$th AP. The aggregated signal at CPU to detect $s_k$ is given by
\begin{equation}
\begin{split}
\label{eq:CPU_sum}
\hat{s}_k = \sum_{l=1}^{L} a_{kl}^* \check{s}_{kl}.
\end{split}
\end{equation}

By substituting~\eqref{eq:local-data-estimate} in~\eqref{eq:CPU_sum}, we can derive that 

\begin{subequations}
\begin{align} 
\hat{s}_k &=\left(\sum_{l=1}^{L} a_{kl}^* \mathbf{v}_{kl}^{H} \mathbf{h}_{kl}\right)s_k \!+ \!  \sum_{l=1}^{L} a_{kl}^* \!\Bigg(\sum\limits_{i=1,i\ne k}^{K}\mathbf{v}_{kl}^{H} \mathbf{h}_{il} s_i\!\!\Bigg) + \mathbf{n}^\prime_{k} \!\!\label{eq:CPU_linearcomb}\!\!\!\! \\
&=\mathbf{a}_{k}^{H}\mathbf{g}_{kk}s_k + \sum\limits_{i=1,i\ne k}^{K}\mathbf{a}_{k}^{H}\mathbf{g}_{ki} s_i + \mathbf{n}^\prime_{k}, \!\!\label{eq:CPU_linearcomb_eff}
\end{align}
\end{subequations} 
 where  $\mathbf{g}_{ki} = [ \mathbf{v}_{k1}^{H} \mathbf{h}_{i1} \, \ldots \, \mathbf{v}_{kL}^{H} \mathbf{h}_{iL}]^{T} \in \mathbb{C}^L$ is  the receive-combined channels between  $k$th UE and each of the APs, $\mathbf{a}_{k} = [ a_{k1} \, \ldots \, a_{kL} ]^{T} \in \mathbb{C}^L$ is the weighting coefficient vector, $\{\mathbf{a}_{k}^{H}\mathbf{g}_{ki}: i=1,\ldots,K\}$ is the set of effective channels, and $\mathbf{n}^\prime_{k} = \sum_{l=1}^{L} a_{kl}^* \mathbf{v}_{kl}^{H}\mathbf{n}_{l}$.

Using the channel statistics at the CPU,  the effective SINR of $k$th UE can be expressed as~\cite{DBLP:journals/corr/abs-1903-10611}

\begin{subequations}
\begin{align}
\label{eq:uplink-instant-SINR-level3}
{SINR}_{k}\! &=\!  \frac{ p_{k} \left.| \mathbf{a}_{k}^{H} \mathbb{E}\{ \mathbf{g}_{kk}\} \right.|^2  }{ \!
 \!\sum\limits_{i=1}^{K} p_{i} \mathbb{E} \{  |\mathbf{a}_{k}^{H}\mathbf{g}_{ki}|^2  \} - p_{k} \left.| \mathbf{a}_{k}^{H} \mathbb{E}\{ \mathbf{g}_{kk}\} \right.|^2 \!+\! \sigma^2 \mathbf{a}_{k}^{H}\mathbf{D}_{k}
 \mathbf{a}_{k}
}\\ 
\label{eq:uplink-rearranged-SINR-new}
&=\!  \frac{ p_{k} \left.| \mathbf{a}_{k}^{H} \mathbb{E}\{ \mathbf{g}_{kk}\} \right.|^2  }{ \! \mathbf{a}_{k}^{H}(\!\sum\limits_{i=1}^{K} p_{i} \mathbb{E} \{  \mathbf{g}_{ki} \mathbf{g}_{ki} ^{H}\} - p_{k} \left. \mathbb{E}\{ \mathbf{g}_{kk}\} \mathbb{E}\{ \mathbf{g}_{kk}\}^{H} \right. \!+\! \sigma^2 \mathbf{D}_{k}
  )\mathbf{a}_{k}
}\\ 
\label{eq:uplink-rearranged-SINR-opt}
      &= \!  \frac{ \mathbf{a}_{k}^{H}  \left.( p_{k} \mathbb{E}\{ \mathbf{g}_{kk}\} (\mathbb{E}\{ \mathbf{g}_{kk}\})^{H}  \right.) \mathbf{a}_{k} }{ \! \mathbf{a}_{k}^{H}(
 \!\sum\limits_{i=1}^{K} p_{i} \mathbb{E} \{  \mathbf{g}_{ki} \mathbf{g}_{ki}^{H}  \} - p_{k} \left. \mathbb{E}\{ \mathbf{g}_{kk}\} \mathbb{E}\{ \mathbf{g}_{kk}\}^{H} \right. \!+\! \sigma^2 \mathbf{D}_{k}
  )\mathbf{a}_{k}
},
\end{align}
\end{subequations}
where $\mathbf{D}_{k}=(\mathbb{E}\{ \|  \mathbf{v}_{k1}\|^2\}, \ldots, \mathbb{E}\{\|  \mathbf{v}_{kL}  \|^2\})\in \mathbb{C}^{L \times L}$ and the expectations are with respect to all sources of randomness. Note that the uplink effective SINR of  $k$th UE can be formulated as a generalized Rayleigh quotient~\cite{massivemimobook}  with respect to $\mathbf{a}_{k}$.

Assuming that UEs transmit with fixed transmit powers, we maximize  generalized Rayleigh quotient in \eqref{eq:uplink-rearranged-SINR-new}. Hence, the optimal weighting coefficient vector of $k$th UE with fixed power constraints is given by~\cite{DBLP:journals/corr/abs-1903-10611}

\begin{equation} \label{eq:LSFD-vector}
\mathbf{a}_{k} =  \left( \sum\limits_{i=1}^{K}
p_{i} \mathbb{E} \{ \mathbf{g}_{ki} \mathbf{g}_{ki}^{H} \} + \sigma^2 \mathbf{D}_{k}
  \right)^{\!\!-1}  \mathbb{E}\{ \mathbf{g}_{kk}\}.
\end{equation}
Aforementioned optimal weighting coefficients lead to the maximum SINR value of $k$th UE with fixed transmit power constraints~\cite{DBLP:journals/corr/abs-1903-10611}

\resizebox{.95\linewidth}{!}{
\begin{minipage}{\linewidth}
\begin{align} \notag
\label{eq:fixed-power}
&{SINR}_{k,max} = p_{k}  \mathbb{E}\{ \mathbf{g}_{kk}^{H}\} \\
& \times \!\! \left( \sum\limits_{i=1}^{K}
p_{i} \mathbb{E} \{ \mathbf{g}_{ki} \mathbf{g}_{ki}^{H} \} \!+\! \sigma^2 \mathbf{D}_{k}
\!-\! p_{k} \mathbb{E}\{ \mathbf{g}_{kk}\} \mathbb{E}\{ \mathbf{g}_{kk}^{H}\}
  \right)^{\!\!-1} \!\!\!\mathbb{E}\{ \mathbf{g}_{kk}\}.
\end{align}
\end{minipage}
 }
 
An achievable SE of $k$th UE  is given by 
\begin{equation} \label{eq:uplink-rate-expression-level3}
\begin{split}
{SE}_{k} = \left( 1 - \frac{\tau_p}{\tau_c} \right) \log_2  \left( 1 + {SINR}_{k}  \right).
\end{split}
\end{equation}
Max-min SINR problem can be formulated such that minimum uplink user SINR is maximized subject to individual transmit power constraint at each UE. This max-min SINR problem can be formulated as follows: 


\begin{align}
 \max_{p_k,\textbf{a}_k} ~~~\min_{k=1,\cdots,K} ~~\text{SINR}_k
\label{p1}
\end{align}
\begin{equation*}
~~~~~~~~~~~\text{subject to}\quad 0 \le p_k \le p_{max}^{(k)}, \quad \forall ~k, \nonumber 
\end{equation*}
\begin{equation*}
~~~~~~~~~~~~~~~~~~~||\textbf{a}_k|| = 1, \quad \forall ~k,
\end{equation*}
where  $p_{max}^{(k)}$ is the maximum transmit power available at $k$th UE.    

\section{Proposed Algorithm}
\label{Sec:prob_formulation}

In this section, we design an optimization methodology based on  alternating solution method  to  maximize the minimum SINR in cell-free  mMIMO system with L-MMSE processing. Problem  (\ref{p1}) is not jointly convex with respect to optimization variables,  $\mathbf{a}_{k}$ and $p_{k}$. Thus, standard convex optimization tools cannot be directly applied to solve problem (\ref{p1}). In the sequel, we propose an iterative algorithm based on GP solution methods and eigenvalue problems to find suboptimal solution for (\ref{p1}), by alternately solving two subproblems, as illustrated in Fig. 2. 
\begin{figure}[t!]
\centering
\includegraphics[width=90mm]{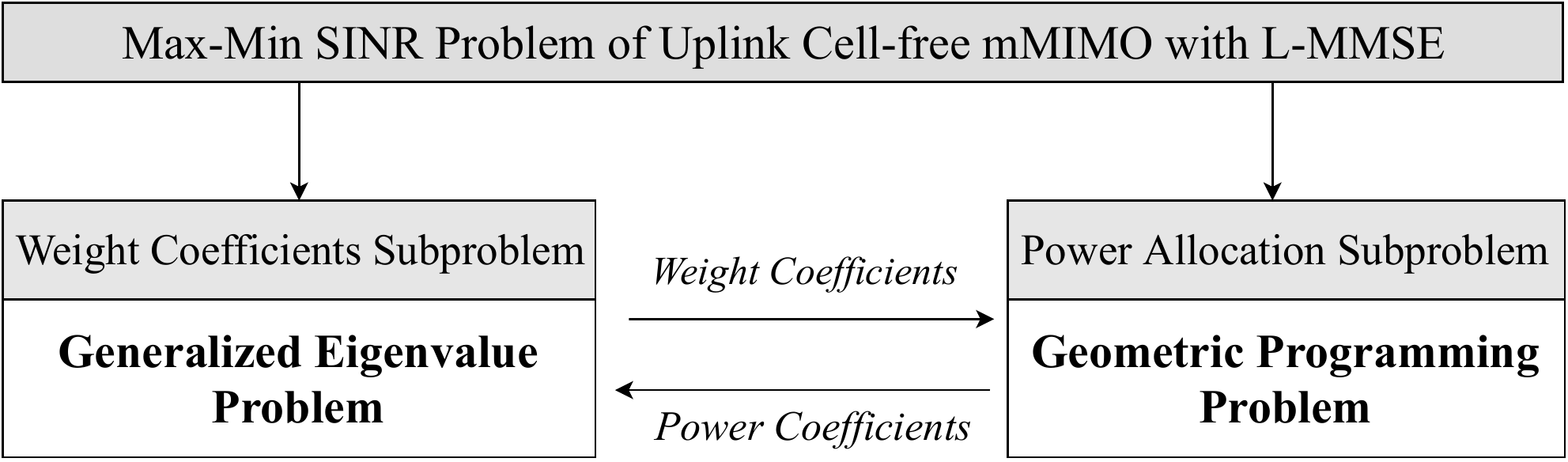}
\caption{Basic idea of the proposed algorithm.}
\label{Fig_Algo}
\end{figure}

\subsection{Weighting Coefficients Design}
\label{sec:subproblem-1}
First, we fix transmit power of $k$th UE and formulate the weighting coefficient subproblem to maximize  uplink SINR  of $k$th user \eqref{eq:uplink-rearranged-SINR-opt}, for all $k$. Then, the  optimal weighting coefficients which corresponds to each UE are obtained by solving corresponding optimization problem in (\ref{p2}).

     

Problem (\ref{p2}) is a generalized eigenvalue problem~\cite{Golub:1996:MC:248979}. The optimal weighting coefficients are obtained from the generalized eigenvector corresponding to the maximum generalized eigenvalue of the matrix pair $\mathbf{A}_{k} = p_k\mathbb{E}\{ \mathbf{g}_{kk}\}\mathbb{E}\{ \mathbf{g}_{kk}\}^H$ and $\mathbf{B}_{k} = \!\sum\limits_{i=1}^{K} p_{i} \mathbb{E} \{  \mathbf{g}_{ki}  \mathbf{g}_{ki}^H  \} - p_{k} \left. \mathbb{E}\{ \mathbf{g}_{kk}\} \mathbb{E}\{ \mathbf{g}_{kk}\}^H \right. \!+\! \sigma^2 \mathbf{D}_{k}$. Note that,  $\mathbf{A}_{k}$ is $\textbf{rank-1}$ matrix.  This makes it possible to find the only nonzero eigenvalue in a closed form. The optimal eigenvector  $\mathbf{a}_{k} =   \mathbf{B}_{k} ^{-1}  \mathbb{E}\{ \mathbf{g}_{kk}\}$ can be obtained by following ~\cite[Lemma B.10]{massivemimobook}. 

\begin{figure*}[b!]
\hrulefill
 \begin{maxi}[2]
	  { \textbf{a}_k}{ \frac{ \mathbf{a}_{k}^{H}  \left( p_{k} \mathbb{E}\{ \mathbf{g}_{kk}\} (\mathbb{E}\{ \mathbf{g}_{kk}\})^{H}  \right.) \mathbf{a}_{k} }{ \! \mathbf{a}_{k}^{H}(
 \!\sum\limits_{i=1}^{K} p_{i} \mathbb{E} \{  \mathbf{g}_{ki} \mathbf{g}_{ki}^{H}  \} - p_{k} \left. \mathbb{E}\{ \mathbf{g}_{kk}\} \mathbb{E}\{ \mathbf{g}_{kk}\}^{H} \right. \!+\! \sigma^2 \mathbf{D}_{k}
  )\mathbf{a}_{k}
} \label{p2}}{}{}
	  \addConstraint{||\textbf{a}_k||}{=1,\quad ~\forall ~k.}
     \end{maxi}

\end{figure*}


\begin{algorithm}[t!]
\caption{}
\textbf{1.} Initialize $\textbf{p}^{(0)}=[p_1^{(0)},p_2^{(0)},\cdots,p_K^{(0)}]$, $i=0$

\textbf{2.} Set $i=i+1$

\textbf{3.} Set $\textbf{p}^{(i)}=\textbf{p}^{(i-1)}$ and find the optimal weighting coefficients $\textbf{a}^{(i)}=[\textbf{a}^{(i)}_1,\textbf{a}^{(i)}_2,\cdots,\textbf{a}^{(i)}_K]$ through solving the generalized eigenvalue Problem  (\ref{p2})

\textbf{4.} Compute $\textbf{p}^{(i)}$ through solving Problem  (\ref{p4})

\textbf{5.} Go back to Step 2 and repeat until required accuracy
\label{al1}
\end{algorithm}

\subsection{ Power Allocation}
\label{sec:subproblem-2}
Next, we formulate the power allocation subproblem by fixing weighting coefficients in master problem (\ref{p1}). The power allocation subproblem is formulated as following max-min SINR optimization problem:
\begin{align}
 \max_{p_k} ~~~\min_{k=1,\cdots,K} ~~\text{SINR}_k
\label{p3}
\end{align}
\begin{equation*}
~~~~~~~~~~~\text{subject to}\quad 0 \le p_k \le p_{max}^{(k)}, \quad \forall ~k. \nonumber 
\end{equation*}


Then, equivalent problem (\ref{p3}) can be obtained by introducing a new slack variable as follows:
\begin{maxi}[2]
{t, p_k}{t~~~~~~~~~~~~ \label{p4}}{}{}
\addConstraint{0 \le p_k \le p_{\text{max}}^{(k)} , \quad}{ \forall ~k}
\addConstraint{\text{SINR}_{k}\ge t , \quad}{   \forall ~k.}
\end{maxi} From \eqref{eq:uplink-rearranged-SINR-opt} the uplink effective SINR of  $k$th UE can be approximated as follows.

\begin{equation} \label{eq:uplink-rearranged-SINR-new-approximated}
{SINR}_{k}\! \approx \!  \frac{ \mathbf{a}_{k}^{H}  \left( p_{k} \mathbb{E}\{ \mathbf{g}_{kk}\} (\mathbb{E}\{ \mathbf{g}_{kk}\})^{H}  \right) \mathbf{a}_{k} }{ \! \mathbf{a}_{k}^{H}(
 \!\sum\limits_{i=1 \ne k}^{K} p_{i} \mathbb{E} \{  \mathbf{g}_{ki} \mathbf{g}_{ki}^{H}  \}  \!+\! \sigma^2 \mathbf{D}_{k}
  )\mathbf{a}_{k}
}
\end{equation}
\textit{Proposition 1}: With the SINR approximation in \eqref{eq:uplink-rearranged-SINR-new-approximated}, problem (\ref{p4}) can be approximated into a GP.
\\
{\textit{Proof:}} See Appendix A.
\\
Hence, with the GP approximation in \eqref{eq:uplink-rearranged-SINR-new-approximated}, problem (\ref{p4})  can be  solved using convex optimization software.

We have now designed two subproblems in \ref{sec:subproblem-1} and  \ref{sec:subproblem-2} which are capable of solving with existing optimization tools. Thus, we now propose an alternating algorithm based on those two subproblems to solve the master problem in (\ref{p1}). The proposed algorithm is summarized in Algorithm \ref{al1}.

\section{Computational Complexity}
\label{sec:complexity}

In this section, we analyse the complexity of Algorithm~1 based on number of floating point operations (flops).
The weighting coefficients design subproblem (\ref{p2}) requires inversion of Hermitian positive definite matrix, which can be efficiently computed using Cholesky factorization~\cite{numericallinearalgebratrefethen97} with $\bigO(\frac{1}{3} KL^{3} )$ flops~\cite{Golub:1996:MC:248979, com_gev}. GP can be solved using interior-point methods as discussed in ~\cite{Boyd2007GP}. Therefore, the power allocation subproblem in (\ref{p4}), which is a GP requires $\bigO( K^{\frac{7}{2}} )$ flops~\cite[Chapter 10.4]{interior_point}. Hence, the overall computational complexity of Algorithm~1 is  $\bigO( K^{\frac{7}{2}} )$  flops per iteration.




\linespread{1.0}

\begin{table}[t]\label{network_parameters}
 \caption{
        Cell-free Massive MIMO network.
    }
\centering
\begin{tabular}{|c|c|}
\hline
Simulation area  &  $1$\,km $ \times\,  1$\,km \\
Bandwidth & $20$\,MHz  \\
Number of APs &  $L$ \\
Number of UEs &  $K$ \\
Number of Antennas per AP &  $N$ \\
UL noise power & $ -96$\,dBm \\
Samples per coherence block & $\tau_c = 200$ \\
Pilot reuse factor & $f$ \\
\hline
\end{tabular}
\label{table1}
\end{table}

\section{Numerical Results and Discussion}
\label{sec:results}
In this section, we provide the numerical evaluation of our proposed optimization methodology~(Algorithm 1) to fixed power optimization scheme~(\ref{eq:fixed-power}). 
The  performance of  cell-free mMIMO system is evaluated by finding the user which corresponds to minimum uplink SE~(\textit{min-user uplink SE}) as the performance metric.

In first subsection, we  discuss the cell-free mMIMO network setup along with its simulation parameters. In second subsection, we present a variety of  simulation results  to evaluate the performance of cell-free mMIMO network model by extending our proposed algorithm. In final subsection, we numerically evaluate the convergence  of Algorithm 1.

\begin{figure}[b!]
\centering
\vspace{-1.5in}
\includegraphics[width=0.6\textwidth]{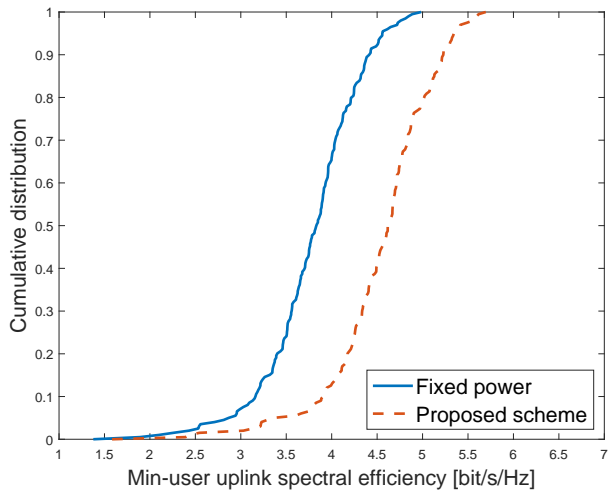}
\vspace{-1.5in}
\caption{The cumulative distribution of  min-user uplink SE with proposed and fixed power schemes ($L=100$, $K=40$, $N=4$, $f=4$,  and $D=1$ km).}
\label{performance}
\end{figure}
\begin{figure}[t!]
\centering
\vspace{-1.5in}
\includegraphics[width=0.6\textwidth]{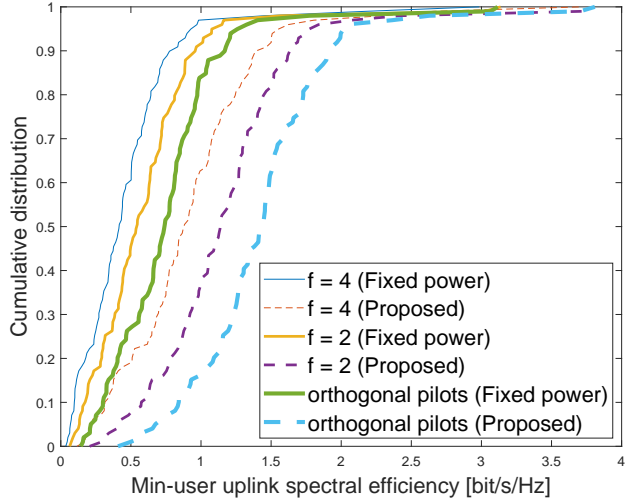}
\vspace{-1.5in}
\caption{The cumulative distribution of  min-user uplink SE with different pilot reuse factors ($L=64$, $K=16$, $N=2$, and $D=1$ km). The dashed curves refer to the proposed Algorithm~1, while the solid curves present the fixed power case.}
\label{reuse_factors}
\end{figure}

\subsection{Parameters and Setup}


We consider a cell-free mMIMO  network with parameters summarized in Table 1. The network consists of uniformly distributed $100$ APs, each with four antennas ($L=100$, $N=4$), covering a square area of $1000\times 1000$ m ($D \times D$). Initially, we  divide simulation area into 4 virtual cells in order to facilitate UE placement. It is assumed that total number of UEs as $K=40$ in which 10 users are randomly dropped  in each cell. For all simulations, we consider communication bandwidth as $20$\,MHz  with a  receiver noise power~($\sigma^2$) of $-96$\,dBm based on~\cite{DBLP:journals/corr/abs-1903-10611}. It is assumed that maximum transmit power of each UE lies  between $90$ mW - $110$ mW and all UEs transmit at their maximum transmit power in the channel estimation phase. Simulations are performed over three cases of pilot reuse factors ($f=1, 4, 8$). For $f=1$ (mutually orthogonal pilot assignment), we assume that $\tau_p =K$,  however $\tau_p$ is further reduced with higher reuse factors.  Eventually, simulation results are averaged over $200$ UE distributions.

As proposed in~\cite{DBLP:journals/corr/abs-1903-10611}, 3GPP urban microcell model is chosen as an appropriate propagation model for cell-free mMIMO system. The carrier frequency is selected as  $2$ GHz. Small scale fading coefficients are generated using correlated Rayleigh fading in which  Gaussian local scattering model with $15^\circ$ angular standard deviation~\cite[Sec.~2.6]{massivemimobook} contributes for the spatial correlation matrix. Moreover, large scale fading coefficients are generated independently as follows~\cite{DBLP:journals/corr/abs-1903-10611}: 
\begin{equation}
\beta_{kl} \,  [\textrm{dB}] = -30.5 - 36.7 \log_{10}\left( \frac{d_{kl}}{1\,\textrm{m}} \right)  + F_{kl},
\end{equation}
where $d_{kl}$ is the distance between UE $k$ and AP $l$ and $F_{kl}$ denotes the shadow fading as $F_{kl}\sim \mathcal{N}(0,4^2)$.

\subsection{Results and Discussions}\label{sec:rnd}

To provide a statistical description of  achievable  SE, we consider empirically the \textit{cumulative distribution}  plots in our simulation results. We thus compare two cumulative distributions of  min-user uplink SE particularly for fixed power optimization scheme and Algorithm 1. We consider the simulation setup with  $100$ of four-antenna ($L=100$, $N=4$) APs and $40$ users ($K=40$). Random pilots are assigned with $f=4$.
Fig. \ref{performance} shows the empirical cumulative distribution plots of the min-user uplink SE for proposed and fixed power schemes.
We can conclude that proposed algorithm outperforms the SE obtained by using the fixed power optimization scheme due to the fact that optimization is performed over both transmit power and weighting coefficients.



Next, we evaluate the cumulative distribution plots for three different cases of pilot assignments; mutually orthogonal pilots, $f=2$, and $f=4$. For this simulation, cell-free mMIMO system is considered with $64$ APs ($L=64$), each with two-antennas ($N=2$) and $16$  users ($K=16$). Fig. \ref{reuse_factors} illustrates the empirical cumulative distribution plots of the min-user uplink SE for the proposed algorithm and fixed power optimization scheme for three different pilot assignments. The  results thus show that even for different pilot assignments, the performance of proposed scheme is higher compared to fixed power case. However, in the case of orthogonal pilot assignment, there is a noticeable increase in the min-user uplink SE compared to non-orthogonal pilot schemes. Among non-orthogonal pilot assignments, $f=2$ case outperforms $f=4$ case due to the fact that channel estimation error increases with pilot reuse. 

\begin{figure}[b!]
\centering
\vspace{-1.5in}
\includegraphics[width=0.7\textwidth]{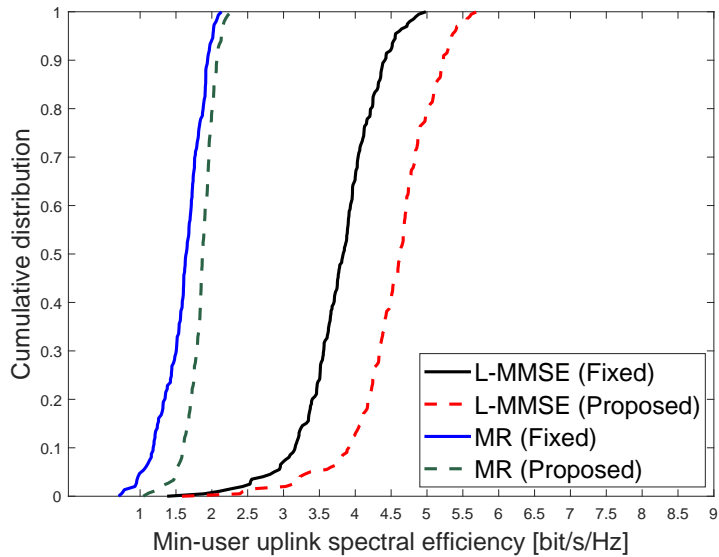}
\vspace{-1.5in}
\caption{The comparison of cumulative distributions of min-user uplink SE using L-MMSE and MR, with random pilots for $L=100$, $K=40$, $N=4$, $f=4$, and $D=1$ km. The dashed curves refer to the proposed Algorithm~1, while the solid curves present the fixed power case.}
\label{performancecomparison}
\end{figure}

 Next, we compare empirical cumulative distribution plots of  min-user uplink SE for two receiver processing techniques; L-MMSE and MR. For this simulation, cell-free mMIMO system is considered with $100$ of four-antenna ($L=100$, $N=4$) APs and $40$ users ($K=40$). Random pilots are assigned with $f=4$. As it is illustrated empirically in  Fig. \ref{performancecomparison},  L-MMSE can outperform MR further after optimizing with our proposed algorithm.

\subsection{Convergence Analysis}\label{sec:conv}
We now numerically investigate the convergence behaviour of our proposed algorithm. The idea is to study ``min-user uplink SE versus iteration number''.

\begin{figure}[t!]
\centering
\vspace{-1.5in}
\includegraphics[width=0.7\textwidth]{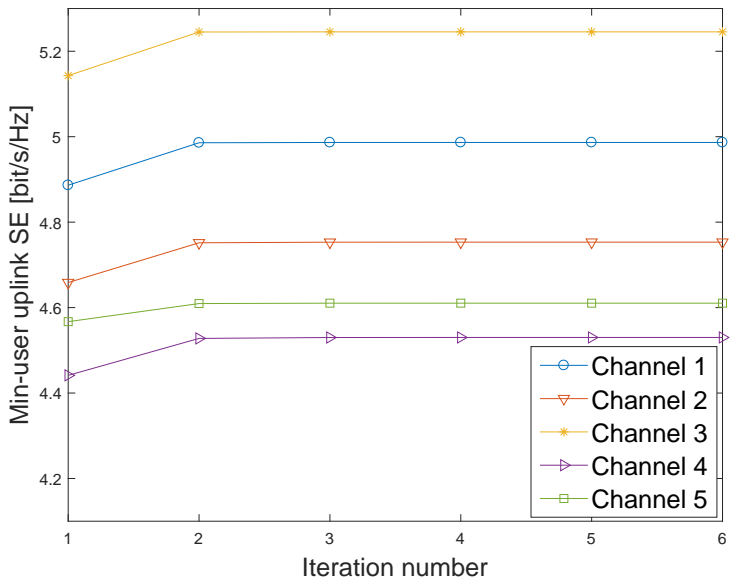}
\vspace{-1.5in}
\caption{The convergence of the proposed algorithm over different set of channel realizations for  $L=100$, $K=40$, $N=4$, $f=4$, and $D=1$ km.}
\label{convergence}
\end{figure}

 In our first simulation setting, we consider only  $5$ different sets of channel realizations. Then, in each set of  channel realizations, we carry out our proposed algorithm for $6$ number of iterations. 
Fig.~\ref{convergence}  shows that min-user uplink SE converges after second iteration. With the results, we can  conclude that the proposed algorithm converges to a suboptimal solution within few iterations. 

\begin{figure}[t!]
\centering
\vspace{-1.5in}
\includegraphics[width=0.7\textwidth]{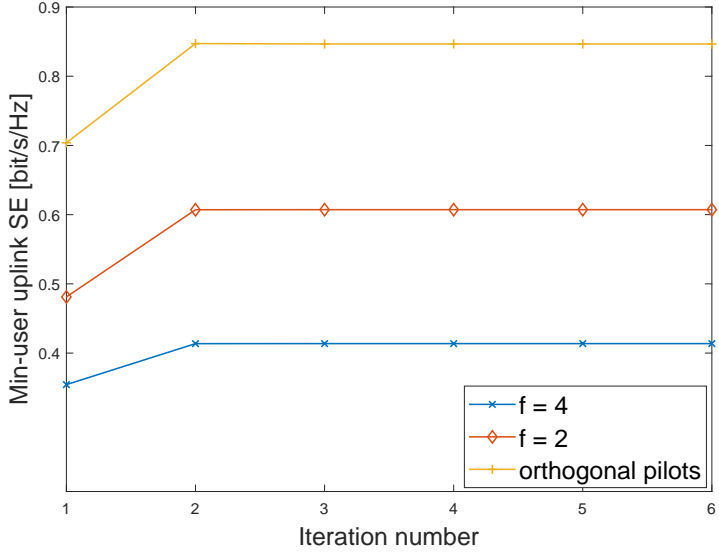}
\vspace{-1.5in}
\caption{The convergence of the proposed algorithm over same set of channel realizations for $L=64$, $K=16$, $N=2$, $D=1$ km, and $f=4,2,1$.}
\label{convergence_multiple}
\end{figure}

Next, we further investigate the convergence of proposed algorithm  for another simulation setup with $64$ APs, each with two antennas, and $16$ users ($L=64$, $N=2$ and $K=16$). In this case, we evaluate the convergence of Algorithm 1 over multiple pilot reuse factors; but for the same set of channel realizations.
Fig. \ref{convergence_multiple} shows the convergence behaviour particularly for  three different pilot reuse factors~($f=1, 4, 8$) when the same set of channel realizations is used. We can thus conclude that the convergence  of proposed algorithm to a suboptimal solution is still guaranteed within few iterations even for the case of different pilot reuse factors. However, as shown in Fig.\ref{convergence_multiple}, orthogonal pilots achieves higher min-user SE compared to non-orthogonal pilot assignments. It is due to the same reasoning as stated  previously for Fig. \ref{reuse_factors} on the effect of  pilot reuse factors for channel estimation.
\\

\section{Conclusions}
\label{sec:conclusion}
We have considered a cell-free mMIMO system which is a promising next generation candidate; particularly for 5G and beyond wireless networks. We  considered the uplink transmission of a cell-free mMIMO network in which local-MMSE combining is employed. To preserve user fairness, the criterion of maximizing the minimum SINR was chosen.  This max-min SINR optimization problem was not jointly convex with respect to the optimization variables; weighting and power coefficients. Therefore, we offered a new optimization methodology to find a suboptimal solution for considered max-min SINR optimization problem by addressing  this non-convexity issue. To cater for the non-convexity, we have   decomposed the original problem into two subproblems. Weighting coefficient subproblem was solved using generalized eigenvalue problem whereas the approximated  version of power coefficient problem was solved by formulating as geometric programming. 

Next, we numerically evaluated the performance of the proposed algorithm with respect to the fixed power optimization scheme where transmit powers were fixed; but weighting coefficients were optimized.
Simulations results empirically  illustrated that the proposed algorithm achieved higher min-user SE  relative to fixed power optimization scheme. Next, we compared the statistical behaviour of min-user SE for L-MMSE and MR receiver processing under both fixed power and proposed schemes. Simulation results showed that L-MMSE performed even better with respect to MR  with our proposed scheme. Moreover, we  numerically investigated the convergence of the proposed alternating algorithm. According to the results, the proposed algorithm required no additional precautions in the initialization, and it converged to a suboptimal solution within few iterations.



\linespread{1.0}

\section*{Appendix A: Proof of Proposition 1}

\begin{IEEEproof} The standard form of GP is defined as follows~\cite{Boyd2007GP}:

\begin{mini}[2]
	  {\textbf{x}}{f_0(\textbf{x})~~~~~~~~~\label{p5}}{}{}
	  \addConstraint{f_i(\textbf{x}) \le 1 ,}{ i= 1,\cdots, m}
	    \addConstraint{g_i(\textbf{x}) = 1 ,}{ i= 1,\cdots, p},
     \end{mini}
where $f_i$ is a posynomial function and $g_i$ is a monomial function. The optimization variables are denoted by $\textbf{x}$ which is an n-dimensional vector.

The form of the SINR constraint in  (\ref{p4}) is not a posynomial function. Therefore, it can be first rewritten and  then approximated into a posynomial function as follows: 

\begin{equation*}
\label{proofexp1}
\frac{\mathbf{a}_k^H\!\left(
 \!\sum\limits_{i=1}^{K} p_{i} \mathbb{E} \{  \mathbf{g}_{ki}\mathbf{g}_{ki}^{H}  \} - p_{k}\left. \mathbb{E}\{ \mathbf{g}_{kk}\} \mathbb{E}\{\mathbf{g}_{kk}\}^H \right. \!+\! \sigma^2 \mathbf{D}_{k}\right)\mathbf{a}_k}{\mathbf{a}_k^H\!\left.( p_{k} \mathbb{E}\{\mathbf{g}_{kk}\} (\mathbb{E}\{\mathbf{g}_{kk}\})^H  \right.)\mathbf{a}_k}\! \le \!\frac{1}{t}.\nonumber 
\end{equation*}
\vspace{3mm}\\
From \eqref{eq:uplink-rearranged-SINR-new-approximated}, SINR constraint can be approximated as follows:

\begin{equation}
    \label{proofexp2}
\frac{\mathbf{a}_k^H\!\left(
 \!\sum\limits_{i=1 \ne k}^{K} p_{i} \mathbb{E} \{  \mathbf{g}_{ki}\mathbf{g}_{ki}^{H} \}  \!+\! \sigma^2 \mathbf{D}_{k}
  \right)\mathbf{a}_k}{\mathbf{a}_k^H\!\left( p_{k} \mathbb{E}\{ \mathbf{g}_{kk}\} (\mathbb{E}\{ \mathbf{g}_{kk}\})^{H}  \right)\mathbf{a}_k}\! \le \!\frac{1}{t}, \\  \forall k.
\end{equation}
With a simple rearrangement, (\ref{proofexp2}) can be converted to following equivalent inequality as follows:

\begin{equation}
p_k^{-1}\left(\sum_{i \ne k}^K\!a_{ki}p_{i}\!+c_k\right)<\dfrac{1}{t}, \\  \forall k,
\label{proofexp3}
\end{equation}

where
\begin{equation*}
a_{ki}=\frac{\mathbf{a}_k^H (\mathbb{E} \{  \mathbf{g}_{ki}\mathbf{g}_{ki}^{H}  \}) \mathbf{a}_k}{\mathbf{a}_k^H\!(\mathbb{E}\{ \mathbf{g}_{kk}\} (\mathbb{E}\{ \mathbf{g}_{kk}\})^{H})\mathbf{a}_k}
\end{equation*}
 ~~~and
\begin{equation*}
c_k=\frac{\mathbf{a}_k^H \textbf{D}_{k} \mathbf{a}_k}{\mathbf{a}_k^H\!(\mathbb{E}\{ \mathbf{g}_{kk}\} (\mathbb{E}\{ \mathbf{g}_{kk}\})^{H})\mathbf{a}_k}.
\end{equation*}

\vspace{2mm}
The left-hand side of (\ref{proofexp3}) is a posynomial function. Both inequality constraint and objective function are in the form of posynomial function. Therefore, the approximated version of the power allocation problem (\ref{p4}) is a standard GP problem as defined in (\ref{p5}). 
\end{IEEEproof}
\bibliographystyle{IEEEbib}
\bibliography{referencelist}
\end{document}